\documentclass[aps,prd,onecolumn,groupedaddress,preprintnumbers,superscriptaddress,nofootinbib]{revtex4}
\usepackage{amsmath,amssymb,latexsym}

\usepackage{graphicx}

\begin{document}

\title{
Constraints on Wrapped DBI Inflation in a Warped Throat
}


\author{
Takeshi Kobayashi
}\email[]{tkobayashi_at_utap.phys.s.u-tokyo.ac.jp}
\affiliation{
Department of Physics, School of Science, The University of Tokyo,
Hongo 7-3-1, Bunkyo-ku, Tokyo 113-0033, Japan
}
\author{
Shinji Mukohyama
}\email[]{mukoyama_at_phys.s.u-tokyo.ac.jp}
\affiliation{
Department of Physics, School of Science, The University of Tokyo,
Hongo 7-3-1, Bunkyo-ku, Tokyo 113-0033, Japan
}
\affiliation{
Research Center for the Early Universe, School of Science, The
University of Tokyo, Hongo 7-3-1, Bunkyo-ku, Tokyo 113-0033, Japan
}
\author{Shunichiro Kinoshita}
\email[]{kinoshita_at_utap.phys.s.u-tokyo.ac.jp}
\affiliation{   
Department of Physics, School of Science, The University of Tokyo,
Hongo 7-3-1, Bunkyo-ku, Tokyo 113-0033, Japan
}

\preprint{UTAP-585} 
\preprint{RESCEU-84/07}

\date{\today}

\begin{abstract}
 We derive constraints on the tensor to scalar ratio and on the
 background charge of the warped throat for DBI inflation driven by
 D$5$- and D$7$-branes wrapped over cycles of the throat. It is shown 
 that the background charge well beyond the known maximal value is
 required in most cases for DBI inflation to generate cosmological
 observables compatible with the WMAP3 data. Most of the results derived in
 this paper are insensitive to the details of the inflaton potential,
 and could be applied to generic warped throats.
\end{abstract}

\pacs{???}
\maketitle

\section{Introduction}
\label{sec:intro}

Numerous attempts have been made to explain inflation from
string theory. Among them, brane inflation in a flux compactified warped
throat \cite{Kachru:2003sx} is one of the most promising ideas. However,
it was soon realized that this model had a serious problem in realizing
slow-roll without the help of fine tuning. Ever since, many interesting
ideas have been put forward to overcome this so-called $\eta$ problem
(\cite{Firouzjahi:2003zy,Berg:2004ek,Cline:2005ty}, etc.). 

On the other hand, an alternative approach to the $\eta$ problem which
gives up slow-roll was suggested in the idea of DBI inflation
\cite{Silverstein:2003hf,Alishahiha:2004eh}.
In this model, a relativistic motion of a D3-brane is considered.
Nevertheless, the potential energy dominates over the kinetic energy
since the latter is suppressed due to the warping of the throat,
which leads to an accelerated expansion of the universe. 

Besides being a remedy for the $\eta$ problem, DBI inflation has a further
exciting property which gathered attention. It has a possibility of
producing signals in the temperature fluctuation in the cosmic microwave
background radiation within the observable range of future detectors.
Especially, it is expected to produce large non-Gaussianity. These
features can be attributed to DBI inflation being a kind of
$k$-inflation with general speed of
sound~\cite{ArmendarizPicon:1999rj,Garriga:1999vw}. 

However, recently it has been pointed out that DBI inflation driven by a 
mobile D$3$-brane might contradict the current WMAP3
data~\cite{Baumann:2006cd,Bean:2007hc,Lidsey:2007gq,Kecskemeti:2006cg}.  
One of the essential points is that one can derive an universal lower bound 
on the tensor to scalar ratio $r$ under the assumption that the
non-Gaussianity is large. Another, more microscopic, point is that $r$
is related to the change of the inflaton $\Delta\phi$ by the so-called Lyth
bound~\cite{Lyth:1996im} (see (\ref{eqn:Lyth-bound}) below). Therefore,
the lower bound of $r$ requires large $\Delta\phi$ over observable
scales. However, since the inflaton field represents the radial position
of the brane, $\Delta\phi$ is restricted by the size of the extra
dimensions. We shall review more details of the argument in
Section~\ref{sec:rev} but the essence is, as mentioned
above, the universal lower bound of $r$ implied by the large
non-Gaussianity and the relation between $\Delta\phi$ and the size of the
extra dimensions. 

In this paper we consider a simple extension of the DBI inflation
model in which the relation between $\Delta\phi$ and the size of the extra
dimensions is modified. We shall consider higher dimensional D-branes wrapped
over cycles of the throat, instead of a simple D$3$-brane. The 
volume of the cycles appears as an overall factor of the 
kinetic term of the inflaton. Thus, properly normalizing the definition
of the inflaton field, the relation between $\Delta\phi$ and the
size of the extra dimensions should be modified. The larger the volume of
the cycles, the larger $\Delta\phi$ for the same size of extra
dimensions. Therefore, it is rather natural to expect that higher
dimensional, wrapped D-branes is one of the simplest scenarios to bypass
the above mentioned problem of DBI inflation. 

For this reason, in this paper, we focus on a D$5$- or D$7$-brane moving 
with a relativistic speed towards the tip of a warped throat, hoping to
find a model which generates large non-Gaussianity and is consistent
with WMAP3 data. Unfortunately, contrary to the hope, in many cases we
find difficulties. We find that DBI inflation with a D$5$-brane requires
a large Euler number of a Calabi--Yau four-fold, exceeding the known
maximal value $\chi\! =\! 1820448$ \cite{Klemm:1996ts}. However, we also
show that a D$7$-brane may be able to excite DBI inflation provided that
the observed CMB scale is produced when the D$7$-brane is in the region
of the throat where the contribution of the NS-NS 2-form potential $B_2$
to the action is substantial. 

The outline of the paper is as follows: In Section~\ref{sec:setup} we
lay out the basic setup. The expressions for the cosmological
observables in DBI inflation are reviewed in Section~\ref{sec:CosmoObs}.
In Section~\ref{sec:rev} we review the constraints on gravitational
waves derived by Lidsey and Huston \cite{Lidsey:2007gq}, and then extend
the discussion to higher dimensional D-branes in
Section~\ref{sec:disc}. We present more stringent bounds focusing on
background charge in Section~\ref{sec:charge}, and we conclude in
Section~\ref{sec:conc}. As an example of a flux compactified warped
throat, the Klebanov--Strassler solution \cite{Klebanov:2000hb,Herzog:2001xk}
is introduced in Appendix~\ref{sec:KS}. The behavior of the inflaton
near the tip of the throat is discussed in Appendix~\ref{sec:tip}. In
Appendix~\ref{sec:Gregion}, we discuss the number of $e$-foldings produced
in the region where $B_2$ can be neglected. Also, a brief discussion on
$\mathrm{det}(G_{kl}-B_{kl})$ is given in Appendix~\ref{sec:det}.

\section{The Basic Setup}
\label{sec:setup}

We assume a warped throat background with its moduli stabilized based on
flux compactification of type IIB string theory. The generic
10-dimensional metric takes the following form: 
%
\begin{equation}
 ds^2 = h^2(\rho)\,  \eta_{\mu\nu}\, dx^{\mu}dx^{\nu}
  + h^{-2}(\rho)\left( d\rho^2 + \rho^2\, g^{(5)}_{mn}\, dx^mdx^n\right),
  \label{metric}
\end{equation}
where $x^{\mu}$ ($\mu = 0,\ 1,\ 2,\ 3$) are the external 4-dimensional
coordinates, $\rho$ is the radial coordinate which decreases as it
approaches the tip of the throat, and $x^m$ ($m=5,\ 6,\ 7,\ 8,\ 9$) are
the internal 5-dimensional angular coordinates. 

In this paper, we investigate conditions that can be imposed upon
{\itshape DBI inflation caused by a D-brane moving with relativistic
speed towards the tip of a warped throat}. Since type IIB string theory
contains stable D$p$-branes with $p$ odd, we focus on D3-, D5-, and
D7-branes and collectively refer to them as D$(3\!+\!2n)$-branes ($n=0,\
1,\ 2$). We assume that a D$(3\!+\!2n)$-brane stretches out along the
external space and also wraps a $2n$-cycle in the
angular directions of the internal space.

DBI inflation is motivated by the Dirac--Born--Infeld (DBI) action, which
takes the following form for a D$(3\!+\!2n)$-brane,
%
\begin{equation}
 S = -T_{3+2n}\int d^{4+2n}\xi\, e^{-\Phi} 
  \sqrt{-\det (G_{AB}-B_{AB})},
\end{equation}
where $G_{AB}$ is the induced metric on the D-brane world-volume,
$B_{AB}$ is the pull-back of the NS-NS 2-form flux $B_2$, $\Phi$ is the
dilaton, and $T_{3+2n}=1/(2\pi)^{3+2n}g_s {\alpha'}^{n+2}$ is the brane
tension. We assume that the dilaton is stabilized to a constant value
and is set to $0$ hereafter.

We take the first four brane coordinates $\xi^{\alpha}$ to coincide
with $x^{\alpha}$ ($\alpha = 0,\cdots,3$) and that the angular position
of the brane in the internal space are functions of the remaining $2n$
brane coordinates $x^m = x^m(\xi^l)$. Then, by further assuming that the
D-brane radial position $\rho$ depends only on $\xi^{\alpha}$, the
induced metric is  
%
\begin{equation}
 G_{AB}\, d\xi^A\ d\xi^B = (h^2 \eta_{\alpha \beta} +
  h^{-2}\partial_{\alpha}\rho\, \partial_{\beta}\rho)
  d\xi^{\alpha}\, d\xi^{\beta} 
  + G_{kl}\, d\xi^k \, d\xi^l, \label{G}
\end{equation}
where 
%
\begin{equation}
 G_{kl} = h^{-2}\rho^2 g^{(5)}_{mn}
  \frac{\partial x^m}{\partial \xi^k}
  \frac{\partial x^n}{\partial \xi^l}.
\end{equation}

For $B_2$, we assume it to have a logarithmic radial dependence and 
have components only along the angular directions. This is a well motivated
property for $B_2$ in order for the dual description to reproduce a
logarithmic flow of couplings found in field theory
\cite{Klebanov:1999rd,Klebanov:2000nc}. However, we postpone the
explicit form of $B_2$ until Section \ref{sec:charge} and here we merely 
state that the legs of $B_2$ is along the angular directions of the
internal space:
%
\begin{equation}
 B_2 = \frac{1}{2}b_{mn}dx^m\wedge dx^n,
\end{equation}
where $b_{mn}$ is antisymmetric in its indices.

Then the DBI action for a D$(3\!+\!2n)$-brane takes the following form,
%
\begin{equation}
 S = - T_{3+2n} \int d^4\xi\, 
  h^4 \sqrt{1 + h^{-4}\eta^{\alpha\beta}
  \partial_{\alpha}\rho\partial_{\beta}\rho}
  \int d^{2n}\xi\, 
  \sqrt{\det(G_{kl}-B_{kl})},
\end{equation}
where
%
\begin{equation}
 B_{kl} = b_{mn}\frac{\partial x^m}{\partial\xi^k}
  \frac{\partial x^n}{\partial\xi^l}.
\end{equation}

Assuming that the physical energy scale associated with the 4-dimensional
universe is much lower than the energy scale of moduli stabilization, we
promote the $4$-dimensional flat metric $\eta_{\mu\nu}dx^{\mu}dx^{\nu}$
to a curved metric $g_{\mu\nu}^{(4)}dx^{\mu}dx^{\nu}$:
%
\begin{equation}
 S = - T_{3+2n} \int d^4\xi\, \sqrt{-g^{(4)}}
  h^4 \sqrt{1 + h^{-4}g^{(4)\alpha\beta}
  \partial_{\alpha}\rho\partial_{\beta}\rho}
  \int d^{2n}\xi\, 
  \sqrt{\det(G_{kl}-B_{kl})}.
  \label{DBIaction}
\end{equation}

Introducing a new variable and a function
%
\begin{eqnarray}
 d\phi & \equiv & T_{3+2n}^{1/2} 
  \left\{
   \int d^{2n}\xi\, \sqrt{\det(G_{kl}-B_{kl})}
   \right\}^{1/2}\, d\rho , \label{dphi} \\
 T(\phi) & \equiv & T_{3+2n}\,  h^4 
  \int d^{2n}\xi\, \sqrt{\det(G_{kl}-B_{kl})}, \label{T}
\end{eqnarray}
the action turns into a simple form 
%
\begin{equation}
 S=-\int d^4\xi \sqrt{-g^{(4)}}\, T(\phi)
  \sqrt{1+g^{(4)\alpha\beta}\partial_{\alpha}\phi\partial_{\beta}\phi/T(\phi)}. 
\end{equation}
It should be noted that the differences in results we derive in this
paper between D3-branes and higher dimensional wrapped D-branes
originate in the normalization factor of the inflaton in (\ref{dphi}).

Further adding a Chern--Simons term and the inflaton potential to the
DBI action coupled to gravity, the full inflaton 
action takes the familiar form
%
\begin{equation}
  S = \int d^4\xi \, \sqrt{-g^{(4)}} 
   \left[ \frac{M_p^2}{2}\mathcal{R}
    - \left\{T(\phi)
   \sqrt{1+g^{(4)\alpha\beta}\partial_{\alpha}\phi\partial_{\beta}\phi/T(\phi)}
   - T(\phi) + V(\phi) \right\}\right], \label{action}
\end{equation}
where $M_p$ is the reduced Planck mass.

Hereafter, we assume that the metric $g^{(4)}_{\alpha\beta}$ is the
physical $4$-dimensional metric which directly couples to matter
fields on the Standard Model brane. In general, from the viewpoint of
the $4$-dimensional effective field theory there is no symmetry argument
to prohibit the possibility that the Standard Model fields may be
coupled to a conformally transformed metric
$\Omega^2g^{(4)}_{\alpha\beta}$ rather than $g^{(4)}_{\alpha\beta}$
itself, where $\Omega$ is a function of the inflaton. However, from the
higher-dimensional point of view, if the Standard Model brane is
geometrically separated from branes responsible for inflation then such
a coupling should be highly suppressed. This is the situation expected
in multi-throat scenarios, where the Standard Model brane is in a
different throat from the inflationary throat. (Further motivations for
considering multi-throat scenarios are 
reviewed in \cite{Cline:2006hu,Chialva:2005zy,Mukohyama:2007ig}.) To be 
more precise, the induced metric on the Standard Model brane is
$g^{(4)}_{\alpha\beta}$ only up to a conformal factor. However, this
conformal factor is essentially independent of the inflaton for the
reason explained above. Thus, this conformal factor can be considered as
a constant at low energy as far as all moduli, including volume and shape
of extra dimensions and the position of the Standard Model brane, are
properly stabilized. By rescaling the unit of reference, one can set the
constant conformal factor to $1$. (It should be noted that this
rescaling does not affect the values of the dimensionless cosmological
observables, such as the ones we introduce in the next section. However,
it does change the local string scale on the Standard Model brane so that a
large hierarchy may be generated a la Randall and
Sundrum~\cite{Randall:1999ee}.) For these reasons, throughout the
present paper we suppose that matter fields on the Standard Model brane
are directly coupled to the metric $g^{(4)}_{\alpha\beta}$.

Recently there have been attempts to study the inflaton potential in detail
for
D3-branes~\cite{Burgess:2006cb,Baumann:2007np,Baumann:2007ah}. Nevertheless,
with our present understanding of string theory, it is fair to say that 
the form of the potential $V(\phi)$ is not well under theoretical
control, let alone the potential for wrapped D-branes. Therefore, in this
paper we seek constraints on DBI inflation without specifying the
form of the potential. In other words, the results of this paper depend
only on the kinetic term of the inflaton action, and are insensitive to
the Chern--Simons term and the potential term. We leave the potential
arbitrary and focus only on the DBI part of the action.

\section{Cosmological Observables}
\label{sec:CosmoObs}

Cosmological observables such as the spectrum of density and tensor
perturbations generated by the action (\ref{action}) have been studied in 
\cite{Alishahiha:2004eh,Garriga:1999vw,Chen:2006nt,Lidsey:2006ia}. 
We now briefly review the expressions of cosmological observables for DBI
inflation. Due to the nontrivial form of the kinetic term in the action
(\ref{action}), DBI inflation can be interpreted as a kind of
$k$-inflation with a general speed of sound 
\cite{ArmendarizPicon:1999rj,Garriga:1999vw}.

By taking the functional derivative of the action with respect to the
$4$-dimensional metric $g^{(4)}_{\alpha\beta}$, we obtain the following
expression for the stress-energy tensor
%
\begin{equation}
 T_{\alpha\beta} = \gamma \partial_{\alpha}\phi\partial_{\beta}\phi
  - g^{(4)}_{\alpha\beta}\left[ T(\phi)(\gamma^{-1}-1)+V(\phi)\right],
\end{equation}
where $\gamma$ is an analog of a Lorentz factor in special relativity
%
\begin{equation}
 \gamma \equiv \frac{1}
  {\sqrt{
  1+g^{(4)\alpha\beta}\partial_{\alpha}\phi\partial_{\beta}\phi
  /T(\phi)}}.
\end{equation}
For the FRW background with a homogeneous $\phi$, the pressure $p$ and
the energy density $\rho$ take the following form 
%
\begin{equation}
 p = T(\phi)(1-\gamma^{-1})-V(\phi), \ \ \  \rho =
  T(\phi)(\gamma-1)+V(\phi), 
\end{equation}
and $\gamma$ is
%
\begin{equation}
 \gamma = \frac{1}{\sqrt{1-\dot{\phi}^2/T(\phi)}} . \label{gamma}
\end{equation}
The speed of sound relevant to inhomogeneous perturbations is given by 
%
\begin{equation}
 c_s = \frac{1}{\gamma}. \label{soundspeed}
\end{equation}

We define the following parameters (often called DBI parameters) in
analogy with the usual slow-roll parameters
%
\begin{eqnarray}
 \tilde{\epsilon} &\equiv & \frac{2 M_p^2}{\gamma}
  \left(\frac{H'}{H}\right)^2, \\
 \tilde{\eta} &\equiv & \frac{2 M_p^2 H''}{\gamma H}, \\
 \tilde{s} &\equiv & 
  \frac{2 M_p^2 \gamma' H'}{\gamma^2 H}, \label{kappa} 
\end{eqnarray}
where $H$ is the Hubble expansion rate. We have adopted the
Hamilton-Jacobi formalism and a prime denotes 
derivatives with respect to the scalar field $\phi$. The absolute values
of these parameters are assumed to be less than one during inflation.

To the lowest order in these parameters, the cosmological observables
are 
%
\begin{eqnarray}
 P_s &=& \frac{1}{8 \pi^2 M_p^2}\frac{H^2}{c_s \tilde{\epsilon}}, \\
 P_t &=& \frac{2}{\pi^2}\frac{H^2}{M_p^2}, \\
 n_s - 1 &=& 2\tilde{\eta} - 4 \tilde{\epsilon} - 2 \tilde{s}, \label{ns} \\
 n_t &=& -2\tilde{\epsilon}, \\
 r &=& 16 c_s \tilde{\epsilon}, \\
 f_{\mathrm{NL}} &=& \frac{1}{3}\left( \frac{1}{c_s^2}-1\right),
  \label{f_NL}
\end{eqnarray}
where $P_s$: scalar perturbation, $P_t$: tensor perturbation, $n_s$: scalar
spectral index, $n_t$: tensor spectral index, $r$: tensor to
scalar ratio, $f_{\mathrm{NL}}$: non-Gaussianity parameter.
Note that the right hand sides should be estimated at the moment of
sound horizon crossing $k c_s=aH$ (although the tensor perturbations
freeze when $k=aH$, the difference is unimportant to lowest order in
the DBI parameters).

Since DBI inflation can be described as a type of brane inflation with a
large Lorentz factor $\gamma$, it is clear from (\ref{soundspeed}) and
(\ref{f_NL}) that DBI inflation generates large non-Gaussianity. In this
paper, we investigate various consistency relations under the assumption
that $|f_{\mathrm{NL}}|$ is large. For a detailed discussion on the
required largeness of $|f_{\mathrm{NL}}|$, see
subsection~\ref{subsec:rev-LB}.

Throughout this paper, the following relation among the observables is
frequently used
%
\begin{equation}
 \frac{\pi^2}{16}r^2 P_s \left(1+\frac{1}{3 f_{\mathrm{NL}}}\right) =
  \frac{T(\phi)}{M_p^4} . \label{TrPs}
\end{equation}
Note that $\phi$ in the right hand side is estimated at the moment the
fluctuation is being produced. Other useful relations are 
%
\begin{equation}
 r = \frac{8}{M_p^2}\left(\frac{d\phi}{d\mathcal{N}}\right)^2
  \label{eqn:pre-Lyth-bound}
\end{equation}
where $\mathcal{N}$ is the number of $e$-folds, and
%
\begin{equation}
 1 - n_s = 4\tilde{\epsilon} + \frac{2\tilde{s}}{1-\gamma^2} - 
  \frac{2M_p^2}{\gamma}\frac{T'H'}{TH}. \label{eqn:identity1}
\end{equation}
In particular, the following corollary of (\ref{eqn:pre-Lyth-bound}) is
called the Lyth bound~\cite{Lyth:1996im}:
%
\begin{equation}
 \left(\frac{\Delta\phi}{M_p}\right)^2 \simeq
  \frac{r}{8}(\Delta\mathcal{N})^2. \label{eqn:Lyth-bound}
\end{equation}

\section{Review of Constraints on DBI Inflation Driven by D$3$-Branes}
\label{sec:rev}

Constraints on gravitational waves for DBI inflation ($f_{\mathrm{NL}}
\gg 1$) with a D3-brane have been derived by Lidsey and Huston
(LH)~\cite{Lidsey:2007gq}, following the work of Baumann and 
McAllister (BM)~\cite{Baumann:2006cd}. We quickly review the discussion
in \cite{Lidsey:2007gq} in this section. Since D3-branes are the focus of
this section, (\ref{dphi}) and (\ref{T}) are simply 
%
\begin{eqnarray}
 d\phi & = & T_3^{1/2} d\rho, \label{dphi-D3}\\
 T & = & T_3 h^4. \label{T-D3}
\end{eqnarray}

\subsection{Lower Bound of $r$}
\label{subsec:rev-LB}

The following relation can be obtained from (\ref{eqn:identity1}),
%
\begin{equation}
 1 - n_s = \frac{r}{4}\sqrt{1+3f_{\mathrm{NL}}}
  - \frac{2\tilde{s}}{3f_{NL}}
  +\frac{\dot{T}}{TH}. \label{eqn:corr-identity1}
\end{equation}

Assuming the following inequality, a lower bound for the tensor to
scalar ratio $r$ can be derived,
%
\begin{equation}
 \dot{T} \leq 0. \label{dotTnegative}
\end{equation}
From (\ref{T-D3}), this is equivalent to $\dot{h}\leq 0$. In other
words, this states that the brane is moving towards the tip of the
throat.

The identity (\ref{eqn:corr-identity1}) combined with the inequality
(\ref{dotTnegative}) gives an inequality relation,
%
\begin{equation}
 \frac{r}{4}\sqrt{1 + 3f_{\mathrm{NL}}} -
  \frac{2\tilde{s}}{3f_{\mathrm{NL}}} \ge 1-n_s \label{lower}.
\end{equation}

We focus on DBI inflation models generating large non-Gaussianity
$|f_{\mathrm{NL}}|$ and a red spectral index $n_s<1$. (The tilt of the
spectrum is preferred to be red by the WMAP3 data. However, if there is
significant negative running in the spectral index, a blue tilted
spectrum is also allowed.)

When $r$ is negligible, then the WMAP3 result $1-n_s > 0.037$ combined
with (\ref{lower}) requires $|\tilde{s}|$ to be large
($|\tilde{s}|>0.05|f_{\mathrm{NL}}|$) and this violates the derivation
of an almost scale invariant power spectrum
(\ref{ns}).\footnote{Furthermore, since higher order terms in DBI
parameters are omitted in deriving the results in
Section~\ref{sec:CosmoObs}, a large $|\tilde{s}|$ will lead to important 
corrections to other cosmological observables as well. Besides, omission
of higher derivative terms of $\phi$ in the DBI action may be
inconsistent when $|\tilde{s}|$ is large.} When $r$ is non-negligible, a
lower bound on $r$ can be obtained from (\ref{lower}), 
%
\begin{equation}
 r \gtrsim \frac{4(1-n_s)}{\sqrt{1+3f_{\mathrm{NL}}}} 
  > \frac{1-n_s}{8} \simeq 0.002. \label{lowerbound}
\end{equation}
The second inequality comes from the WMAP3 limit 
$|f_{\mathrm{NL}}|<300$~\cite{Spergel:2006hy,Creminelli:2006rz}. The far
right hand side is obtained by substituting the WMAP3 best-fit value
$1-n_s \simeq 0.013$.

We should remark that the second term of the left hand side of
(\ref{lower}) was ignored in deriving the first inequality of
(\ref{lowerbound}). This procedure is valid under a small
$\tilde{s}$ and a large $f_{\mathrm{NL}}$. For example, when
$|\tilde{s}|\lesssim 0.1$, $|f_{\mathrm{NL}}|\gtrsim 20$ is
sufficient.

\subsection{Upper Bound of $r$}

Since the $4$-dimensional reduced Planck mass $M_p\equiv (8\pi
G)^{-1/2}$ is given by 
%
\begin{equation}
 M_p^2 = \frac{2}{(2\pi)^7 g_s^2 \alpha '^4}
  \int d\rho \, \mathrm{Vol}(X_5) \frac{\rho^5}{h^4(\rho)},
\end{equation}
it is convenient to define the warped volume of extra dimensions as 
%
\begin{equation}
 V_6 \equiv \int d\rho \, \mathrm{Vol}(X_5) \frac{\rho^5}{h^4(\rho)}.
\end{equation}
Here, $\mathrm{Vol}(X_5)$ is the dimensionless volume of the unit-radius
$5$-dimensional base space ($X_5$) of the throat. Generically, we expect
$\mathrm{Vol}(X_5)$ to be $O(1)\times \pi^3$
(e.g. $\mathrm{Vol}(S^5)=\pi ^3$ for a 5-sphere,
$\mathrm{Vol}(T^{1,1})=\frac{16}{27}\pi ^3$ for a Klebanov--Strassler
(KS) throat which is discussed in Appendix \ref{sec:KS}).

The two inequalities used to derive the upper bound of $r$ are the
following: 
%
\begin{eqnarray}
 \rho_* & > & \Delta \rho,
  \label{i}\\ 
 V_6  & > & \Delta V_6, \label{ii}
\end{eqnarray}
where the subscript ``$*$'' denotes the quantity to be estimated at the
moment the CMB-scale fluctuation is produced, $\Delta \rho$ denotes the
change of the D$3$-brane radial position when the observable scales are
generated, and $\Delta V_6$ is a fraction of the warped volume 
of the throat corresponding to the radial variation $\Delta\rho$. The
validity of the inequalities (\ref{i}) and (\ref{ii}) is clear for DBI
inflation driven by a D-brane moving toward the tip of the throat.

Since $\Delta \rho$ corresponds to no more than $\Delta \mathcal{N}
\simeq 4$ $e$-foldings of inflationary expansion (which is equivalent to
the range $2 \leq l < 100$), $\Delta \rho$ is expected to be a narrow
range in the radial dimension. Hence we adopt the following approximate
expression for $\Delta V_6$:
%
\begin{equation}
 \Delta V_6 \simeq 
\mathrm{Vol}(X_5)\frac{\rho_*^5}{h_*^4}\Delta\rho, \label{DelV}
\end{equation}
where $h_* \equiv h(\phi _*)$.

From (\ref{ii}) and (\ref{DelV}), we obtain
%
\begin{equation}
 \frac{1}{M_p^2} =  \frac{(2\pi)^7 g_s^2 {\alpha'}^4}{2 V_6}
  < \frac{(2\pi)^7 g_s^2 {\alpha'}^4}{2 \Delta V_6}
  \simeq 
  \frac{(2\pi)^7g_s^2{\alpha'}^4h_*^4}
  {2\mathrm{Vol}(X_5)\rho_*^5\Delta\rho}.  \label{D3-1}
\end{equation}
This can be converted further with the use of (\ref{i}),
%
\begin{equation}
 \frac{1}{M_p^2} <
  \frac{(2\pi)^7 g_s^2 {\alpha'}^4 h_*^4}
  {2\mathrm{Vol}(X_5) (\Delta \rho)^6}.
\label{D3-eq}
\end{equation}

From the Lyth bound (\ref{eqn:Lyth-bound}), together with (\ref{TrPs}),
(\ref{dphi-D3}), (\ref{T-D3}), and (\ref{D3-eq}), an upper bound for $r$
can be obtained
%
\begin{equation}
 r < \frac{2^5 \pi^3}{(\Delta \mathcal{N})^6 \mathrm{Vol}(X_5)}
  P_s \biggl(1 + \frac{1}{3 f_{\mathrm{NL}}}\biggr). \label{D3-upper}
\end{equation}

Taking $P_s = 2.5 \times 10^{-9}$ (WMAP3 \cite{Spergel:2006hy}
normalization), $\mathrm{Vol} (X_5) = \pi ^3$, $\Delta \mathcal{N} = 1$
(the most optimistic estimate for the minimum number of $e$-foldings that
can be probed by observation), and ignoring the $f_{\mathrm{NL}}^{-1}$
term since we have assumed $|f_{\mathrm{NL}}|$ to be large, the upper bound 
becomes $r < 10^{-7}$. This obviously contradicts with the lower bound
derived in the previous subsection. (Note that this upper bound is valid 
even if the D$3$-brane is moving away from the tip of the throat, as
long as the inequalities (\ref{i}) and (\ref{ii}) hold.)
Therefore, DBI inflation driven by a D$3$-brane in relativistic
motion (leading to a large $|f_{\mathrm{NL}}|$) always contradicts
current observations. In other words, the above results predict low
velocity of the D-brane and small $|f_{\mathrm{NL}}|$ for inflation
driven by a D$3$-brane in a warped throat, which makes the model
indistinguishable from ordinary slow-roll inflation models.

\subsection{Note on the Difference Between BM and LH}
\label{subsec:diff}

In this section, we have briefly reviewed the constraints on
gravitational waves investigated by LH. Before ending this section, we
should point out some of the main differences between the approaches
taken by BM~\cite{Baumann:2006cd} and LH~\cite{Lidsey:2007gq}.

The first is the derivation of the universal lower bound on $r$ by LH,
which is reviewed in subsection~\ref{subsec:rev-LB}. LH combined this
bound with other constraints derived by BM.

Another difference is that BM considers the total variation of the
D3-brane radial position {\itshape throughout inflation}, which leads to the
introduction of the effective number of $e$-foldings
\begin{equation}
 \mathcal{N}_{\mathrm{eff}}\equiv \int
  ^{\mathcal{N}_{\mathrm{end}}}_0 d\mathcal{N}
  \biggl(\frac{r}{r_*}\biggr)^{1/2}.
\end{equation}
Since it is difficult to estimate the values of cosmological observables
on scales we haven't observed, some assumptions need to be imposed in
order to have a quantitative discussion on
$\mathcal{N}_{\mathrm{eff}}$. In contrast, LH only make use of the
variation of the D3-brane position $\Delta \rho$ and number of
$e$-foldings $\Delta \mathcal{N}$ while the observable scales are
generated, as can be seen from (\ref{i}) and (\ref{ii}). Hence the
approach taken by LH provides more conservative bounds which can be
applied to general cases.

Furthermore, the constraints by LH apply to the case in which
D-branes are moving relativistically, since their derivation rely on
the assumption that $|f_{\mathrm{NL}}|$ is large. On the other hand, the
bounds by BM also apply in the slow roll limit.  

It should also be noted that the results of LH are insensitive to
the details of the throat geometry and the inflaton potential. The
results are directly related to cosmological observables in order to
derive constraints. Meanwhile, BM consider an explicit case in which the
geometry of the throat is $AdS_5 \times X_5$ and the inflaton potential
is quadratic $V(\phi) = \frac{1}{2}m^2\phi^2$. This procedure enables
detailed arguments involving microscopic string theory inputs. As can be
seen in \cite{Baumann:2006cd}, rather stringent bounds on the
background flux can be obtained when the inflaton potential consists
only of a quadratic term.

Throughout this paper we generalize the approach taken by LH and do not
fix the inflaton potential to any form. (For the warp factor, a throat
with $AdS_5 \times X_5$ geometry is considered as an example in
Section~\ref{sec:charge}.)

\section{Extension to D$5$- and D$7$-Branes}
\label{sec:disc}

We now consider higher dimensional D-branes and extend the bounds
on $r$ derived in the previous section for a D$3$-brane to
the case of generic D$(3+2n)$-branes. However, derivation of upper
bounds of $r$ is complicated and subtle for D$5$-
and D$7$-branes. Thus, to make arguments simpler, in this paper we
consider two extreme cases in which $G_{kl}$ or $B_{kl}$ is dominant
over the other in $\mathrm{det}(G_{kl}-B_{kl})$.

In this section, we derive constraints without specifying the
explicit forms of the warp factor and the overall $\rho$-dependence of
the $B_2$ potential. While we derive a general lower bound of $r$ in
subsection~\ref{subsec:gelobr}, the upper bound derived in
subsection~\ref{subsec:rmax-Gregion} holds only in
the region of the warped throat where the effect of $B_2$
can be ignored. There we find that the upper bound of $r$ relaxes
significantly, due to the change in the relation between the
inflaton and the D-brane radial position. It is shown that there is a
regime of $r$ consistent with both the lower and upper bounds.

In the next section we specify the warp factor and the $B_2$ potential
and provide a more complete and stringent discussion on the constraints,
taking into account the tadpole condition and the known maximal value of
the Euler number of a Calabi--Yau four-fold.

\subsection{General Lower Bound of $r$}
\label{subsec:gelobr}

The derivation of the lower bound of the tensor to scalar ratio $r$ in
the previous section relies only on the inequality (\ref{dotTnegative}). 
In this section, however, for the definition of $\phi$ and $T$ we now
have (\ref{dphi}) and (\ref{T}) instead of (\ref{dphi-D3}) and
(\ref{T-D3}). Thus, in this subsection we will derive the lower bound
on $r$ by showing the inequality (\ref{dotTnegative}) for (\ref{dphi})
and (\ref{T}) with $n=1,2$. We assume that the brane is moving towards
the tip of the throat, i.e. $\dot{\rho}<0$. Thus, the inequality
(\ref{dotTnegative}) is equivalent to 
%
\begin{equation}
 \frac{d}{d\rho}
  \left\{ \int d^{2n}\xi
   \sqrt{h^8 \det(G_{kl}-B_{kl})} \right\} \ge 0.\label{ddrho}
\end{equation}

Let us take the angular brane coordinates to
diagonalize $G_{kl}$. If the adequate gauge cannot be chosen throughout
the $2n$-cycle, then we divide the wrapped cycle into patches 
on which proper coordinates can be chosen, and sum up. It is evident
that $G_{kl}$ is a Riemannian metric and, thus, the diagonal components
$G_{ll}$ are positive. Now, we rewrite $G_{kl}$ as
\begin{equation}
 G_{kl} = 
  \frac{\rho^2}{h^2}\mathrm{diag}(\mathcal{G}_5,\cdots,\mathcal{G}_{4+2n}),
 \label{eqn:amplitudeG}
\end{equation}
where $\mathcal{G}_k$ ($>0$) are assumed to satisfy
$\frac{d}{d\rho}(\rho^2\mathcal{G}_k)\geq 0$. (For example, this
inequality is trivially satisfied if $g^{(5)}_{mn}\frac{\partial x^m}{\partial
\xi^k}\frac{\partial x^n}{\partial \xi^l}$ are independent of $\rho$.)
We note that throughout this paper, we focus on a throat with its warp
factor obeying 
\begin{equation}
 \frac{dh}{d\rho} \ge 0. 
\end{equation}

For $B_{kl}$, we assume that it can be decomposed into the following form,
%
\begin{equation}
  B_{kl} = B\mathcal{B}_{kl}, \quad \textrm{with} \quad \frac{dB}{d\rho}
   \ge 0.  \label{eqn:amplitudeB}
\end{equation}
Here, $B$ is a function of $\rho$, while $\mathcal{B}_{kl}$ depends only
on the angular brane coordinates. (For example, the fields can take the
form of (\ref{eqn:amplitudeG}) and (\ref{eqn:amplitudeB})
in the region away from the tip in the KS solution, as can
be seen from (\ref{htildeh}) and (\ref{large-B2}).)

For a D$5$-brane ($n=1$), we obtain
%
\begin{equation}
 h^8 \det(G_{kl}-B_{kl})
  = h^4\rho^4 \mathcal{G}_5\mathcal{G}_6 + h^8B^2 (\mathcal{B}_{56})^2.
\end{equation}
Since each term takes the form of a product of non-decreasing functions
of $\rho$, (\ref{ddrho}) is obvious.

For the case of a D$7$-brane ($n=2$), from the discussion in Appendix
\ref{sec:det}, the following can be obtained,
%
\begin{eqnarray}
 h^8 \det(G_{kl}-B_{kl}) & = &
  \rho^8\mathcal{G}_5\mathcal{G}_6\mathcal{G}_7\mathcal{G}_8
  + h^4\rho^4B^2
  \left[
   \mathcal{G}_5\mathcal{G}_6(\mathcal{B}_{78})^2+ 
   \mathcal{G}_5\mathcal{G}_7(\mathcal{B}_{68})^2+ 
   \right.  \nonumber\\
 & & 
  \left.
   \mathcal{G}_5\mathcal{G}_8(\mathcal{B}_{67})^2+ 
   \mathcal{G}_6\mathcal{G}_7(\mathcal{B}_{58})^2+ 
   \mathcal{G}_6\mathcal{G}_8(\mathcal{B}_{57})^2+ 
   \mathcal{G}_7\mathcal{G}_8(\mathcal{B}_{56})^2
		       \right]  \nonumber\\
 & & + h^8B^4(\mathcal{B}_{58}\mathcal{B}_{67} -
  \mathcal{B}_{57}\mathcal{B}_{68} + \mathcal{B}_{56}\mathcal{B}_{78})^2.
\end{eqnarray}
Similarly, (\ref{ddrho}) is clear.

Hence the condition (\ref{ddrho}) is verified for both D$5$- and
D$7$-branes, which leads to the lower bound
(\ref{lowerbound}). 

Before closing this subsection, we should remark that near the tip of a
deformed conifold, $B_{kl}$ does not take the form of
(\ref{eqn:amplitudeB}). Nevertheless, (\ref{ddrho}) and the lower bound
(\ref{lowerbound}) may still hold. We show this in
Appendix~\ref{sec:tip} through the example of the KS solution.

\subsection{Upper Bound of $r$ in $G_{kl}$ Dominant Region}
\label{subsec:rmax-Gregion}

As the D$5$- or D$7$-brane moves toward the tip of the throat, the
contributions of $G_{kl}$ and $B_{kl}$ to $\det(G_{kl}-B_{kl})$
change. For the case of the KS solution, initially $B_{kl}$ is dominant
over $G_{kl}$, and then $G_{kl}$ becomes dominant in the region closer
to the tip (for a detailed discussion, see Appendix~\ref{sec:Gregion}).

As already stated in the beginning of this section, in this subsection
we seek an upper bound of $r$ without specifying the form of the warp
factor $h$. This is possible if we can neglect $B_{kl}$
compared with $G_{kl}$. This is equivalent to restricting our 
consideration to the region of the throat where $G_{kl}$ is dominant over
$B_{kl}$. The other extreme case, i.e. the $B_{kl}$ dominant region,
will be considered in the next section by using the explicit form of the
warp factor and the NS-NS $2$-form.

Ignoring $B_{kl}$, the integral term in (\ref{dphi}) or (\ref{T})
represents the wrapped volume. Introducing 
%
\begin{equation}
 v_{2n} \equiv \int d^{2n}\xi
  \sqrt{\mathrm{det}\biggl(g_{mn}^{(5)}\frac{\partial x^m}{\partial
  \xi^k}\frac{\partial x^n}{\partial \xi^l} \biggr)},
\end{equation}
which is the unit-radius dimensionless volume of the $2n$-cycle,
(\ref{dphi}) and (\ref{T}) transform to 
%
\begin{eqnarray}
 d\phi & = & T_{3+2n}^{1/2} v_{2n}^{1/2}
  \left(\frac{\rho}{h}\right)^n\, d\rho, \label{G-dphi}\\
 T(\phi) & = & T_{3+2n} v_{2n} h^4 \left( \frac{\rho^2}{h^2}\right)^n.
  \label{G-T}
\end{eqnarray}

Now we combine two inequality relations (\ref{i}) and (\ref{D3-1}) to
obtain an upper bound on $r$. We transform (\ref{i}) into the form
\begin{equation}
 \left(\frac{1}{\rho_*}\right)^{\frac{2\kappa -3n +10}{n+2}} <
  \left(\frac{1}{\Delta \rho}\right)^{\frac{2\kappa -3n +10}{n+2}}.
  \label{new} 
\end{equation}
Here we have introduced an arbitrary parameter $\kappa$ which satisfies
$\frac{2\kappa-3n+10}{n+2}>0$ (hence $\kappa$ can be taken as any
nonnegative number). Later on, we will fix $\kappa$ to an appropriate
value in order to derive the most effective bounds. The combination of
$\kappa$ and $n$ in (\ref{new}) is chosen so that the final inequality
expression (\ref{6}) will contain an equal number of $\rho_*$ and
$1/h_*$.  

Employing the inequality relation (\ref{new}) on the far right hand side
of (\ref{D3-1}), we obtain
%
\begin{equation}
 \frac{1}{M_p^2}<
  \frac{(2\pi )^7\, g_s^2\, {\alpha'}^4}{2\, \mathrm{Vol}(X_5)} \, 
  \frac{h_*^4}{\rho_*^{5-\frac{2\kappa -3n +10}{n+2}} \, \Delta
  \rho} \left(\frac{1}{\rho_*}\right)^{\frac{2\kappa -3n +10}{n+2}} <
  \frac{(2\pi )^7\, g_s^2\, \alpha '^4}{2\, \mathrm{Vol}(X_5)} \, 
  \frac{h_*^4}{\rho_*^{5-\frac{2\kappa -3n +10}{n+2}} \, \Delta
  \rho} \left( \frac{1}{\Delta \rho}\right)^{\frac{2\kappa -3n +10}{n+2}}.
  \label{5}
\end{equation}

From the Lyth Bound (\ref{eqn:Lyth-bound}) and (\ref{G-dphi}),
%
\begin{equation}
 \Delta\rho \simeq \frac{\Delta\mathcal{N}\, M_p\, r^{1/2}}{2^{3/2}\,
  T_{3+2n}^{1/2}\, v_{2n}^{1/2}} \biggl(\frac{h_*}{\rho_*}\biggr)^n.
 \label{3}
\end{equation}

From (\ref{TrPs}) and (\ref{G-T}),
%
\begin{equation}
 r = \frac{4\, T_{3+2n}^{1/2}\, v_{2n}^{1/2}}{\pi \, P_s^{1/2}\, 
  (1+\frac{1}{3f_{\mathrm{NL}}})^{1/2}}\, 
  \left(\frac{h_*}{M_p}\right)^2 \left(\frac{\rho_*}{h_*}\right)^n. 
  \label{2}
\end{equation}

After cancelling out $\Delta \rho$ from the far right hand side of
(\ref{5}) with the use of (\ref{3}), and then cancelling out $M_p$ with
the help of (\ref{2}), an upper bound for $r$ with the following form
can be derived: 
%
\begin{equation}
 r < 
  \frac{(2\pi )^7\, g_s^2\, {\alpha'}^4}{2\,
  \mathrm{Vol}(X_5)}\,
  \left(\frac{4\, T_{3+2n}^{1/2}\, v_{2n}^{1/2}}{\pi \, P_s^{1/2}\, 
   (1+\frac{1}{3f_{\mathrm{NL}}})^{1/2}}
  \right)^{\frac{-\kappa +2n -4}{n+2}}\, 
  \left(\frac{8\,  T_{3+2n}\, v_{2n}}{(\Delta \mathcal{N})^2}
   \right)^{\frac{\kappa -n +6}{n+2}}\, 
  \left(\frac{\rho_*}{h_*}\right)^{\kappa}. \label{6}
\end{equation}
It can be shown that this equation also applies to the case of $n=0$
(when $n=\kappa=0$, (\ref{6}) turns out to be (\ref{D3-upper})). The above
form of the upper bound containing equal numbers of $\rho_*$ and $1/h_*$
is advantageous, as we will see in the subsequent discussions.

In order to obtain a bound without specifying a concrete form of the
warp factor $h(\rho)$, we take $\kappa$ to be $0$. Again we substitute
the following values, which lead to the most optimistic upper bounds (as
is explained below (\ref{D3-upper})):
$P_s = 2.5 \times 10^{-9}$, $\mathrm{Vol}(X_5)=\pi^3$, 
$\Delta\mathcal{N}=1$, and ignoring the $f_{\mathrm{NL}}$ term, we
obtain the following upper bounds of $r$
%
\begin{eqnarray}
 \mathrm{(D5)} & \quad &  
  r < \frac{2^3\, \pi\, g_s^{2/3}\, v_2^{4/3}}{\mathrm{Vol}(X_5)\, 
  (\Delta \mathcal{N})^{10/3}}
  \left\{ P_s \left(1+\frac{1}{3 f_{\mathrm{NL}}}\right)\right\}^{1/3}
  \simeq 1.1 \times 10^{-3}\,  g_s^{2/3}\, v_2^{4/3}, \\
 \mathrm{(D7)} & \quad &  
  r < \frac{4 \, g_s \, v_4}{\mathrm{Vol}(X_5)\, 
  (\Delta\mathcal{N})^2} \simeq 0.13 \, g_s\, v_4.
\end{eqnarray}
We expect $v_{2n}$ to obey $v_2 \sim 4\pi$ (which is the value for a
2-sphere) and $v_4 \sim \frac{8}{3}\pi^2$ (the value for a
4-sphere). Hence, as long as the string coupling constant $g_s$ is
larger than $\mathcal{O}(10^{-2})$ for a D5- and $\mathcal{O}(10^{-4})$
for a D7-brane, the upper bound for $r$ is compatible with the lower bound.

One may expect to obtain more stringent bounds by assigning some number
other than $0$ to $\kappa$ in (\ref{6}). However, in order to do so, the
explicit form of the warp factor is needed. In view of this, in the
following section, we focus on a warped throat with an AdS geometry
and derive more severe constraints.

\section{More Stringent Bounds}
\label{sec:charge}

In the previous section we found that the upper bounds of $r$ for D$5$-
and D$7$-branes are significantly weaker than that for a D$3$-brane and
are compatible with the lower bound. However, in this section we will
find that a large number of background charge is needed. 
In many cases this requires too large an Euler number of a
Calabi--Yau four-fold, well beyond the known maximal value. 

To make our arguments concrete, in this section we assume that the
throat is approximately $\mathrm{AdS}_5\!\times\! X_5$ away from the
tip. Though the throat geometry deviates from the AdS geometry as one
approaches the tip for the case of a deformed conifold, we focus on the
AdS region of the warped throat and estimate various constraints. The
validity of this procedure is shown in Appendix~\ref{sec:tip}, where it
is shown through the example of a KS throat that the observed CMB scale
is generated away from the tip in DBI inflation.

The warp factor in the AdS region is $h(\rho) = \rho /R$ with the
AdS radius
%
\begin{equation}
 \frac{\rho}{h(\rho)} = R
  = \left( \frac{2^2 \pi^4 g_s \alpha'^2 N}{\mathrm{Vol}(X_5)}
			     \right)^{1/4}, \label{R}
\end{equation}
where $N$ is the background number of charges~\cite{Gubser:1998vd}.

The warp factor at the tip of the throat is characterized by the
integers $M$ and $K$ associated with R-R and NS-NS fluxes respectively,
as~\cite{Giddings:2001yu}
%
\begin{equation}
 h(0) \sim \exp \left( -\frac{2\pi K}{3 g_s M}\right). \label{h0}
\end{equation}

As we briefly mentioned in Section \ref{sec:setup}, in many cases, 
the NS-NS 2-form potential $B_2$ depends on the radial coordinate
logarithmically in the AdS region of the throat and have legs along the
angular directions of the internal space. Therefore let us consider the
case of $B_2$ taking the following
form~\cite{Klebanov:1999rd,Klebanov:2000nc}:
%
\begin{equation}
 B_{kl} = B \mathcal{B}_{kl}, \quad
  B = g_s M \alpha ' \ln\left(\frac{\rho}{\rho_b}\right).
\end{equation}
where $M$ is an integer associated with R-R 3-form $F_3$, and
$\mathcal{B}_{kl}$ is independent of $\rho$
as in (\ref{eqn:amplitudeB}). For example, as can be seen
in (\ref{large-B2}), the KS solution in the large $\tau$ region indeed
has $B_{kl}$ of this form. We assume $\rho_b$ to be roughly the place of
the boundary between the AdS region and the region near the tip where
the warp factor is nearly constant. Then $\rho_b$ takes the following
form: 
%
\begin{equation}
 \rho_b \sim (g_s M \alpha ')^{1/2} 
  \exp \left( -\frac{2 \pi K}{3 g_s M}\right), \label{rho-b}
\end{equation}
where $K$ is an integer associated with the NS-NS 3-form $H_3$. (For the
value of $\rho_b$ in the KS solution, see (\ref{rhob}).)

The product of $M$ and $K$ produces the net background charge $N$, and
from the tadpole condition it is related to the topology of a Calabi--Yau
four-fold:
%
\begin{equation}
 KM = N = \frac{\chi}{24} \le 75852, \label{euler}
\end{equation}
where $\chi$ is the Euler characteristic of a CY four-fold and the
inequality on the right hand side comes from the known
maximal value $\chi = 1820448$ \cite{Klemm:1996ts}.

As already stated in the previous section, derivation of the upper bounds of
$r$ for D$5$- and D$7$- branes is rather complex because of the presence
of the determinant of $(G_{kl}-B_{kl})$ in the action. Therefore we 
consider two extreme cases in which $G_{kl}$ or $B_{kl}$ is
dominant over the other. In the following subsections, we consider each
case separately.

Moreover, in Appendix~\ref{sec:Gregion}, the number of
$e$-foldings generated in the $G_{kl}$ dominant region is roughly
estimated and we discuss the place where the CMB scale is produced.

\subsection{Lower Bound of $N$ in $G_{kl}$ Dominant Region}

In this subsection we consider the case in which $G_{kl}$ is dominant
over $B_{kl}$. As discussed in Appendix~\ref{sec:Gregion}, this  
corresponds to the region relatively closer to the tip.

We have already considered this region in
subsection~\ref{subsec:rmax-Gregion} and obtained the upper bound of $r$
(\ref{6}). With the use of (\ref{R}), we now reinterpret (\ref{6}) as a
lower bound of the background charge $N$: 
%
\begin{equation}
 N > 
  \frac{\mathrm{Vol}(X_5) (\Delta \mathcal{N})^{\frac{8}{2+n}}}
  {2^{\frac{2(1-n)}{2+n}} \pi^{\frac{6}{2+n}} g_s^{\frac{n}{2+n}}
  v_{2n}^{\frac{2}{2+n}}P_s^{\frac{2}{2+n}}
  (1+\frac{1}{3f_{\mathrm{NL}}})^{\frac{2}{2+n}}}
  \left[
   \frac{\mathrm{Vol}(X_5) (\Delta \mathcal{N})^{\frac{2(6-n)}{2+n}}r}
   {2^\frac{10-n}{2+n} \pi^{\frac{3(2-n)}{2+n}}
   g_s^{\frac{2n}{2+n}} v_{2n}^{\frac{4}{2+n}}
   P_s^{\frac{2-n}{2+n}}(1+\frac{1}{3f_{\mathrm{NL}}})^{\frac{2-n}{2+n}}}
   \right]^{4/\kappa}, \label{G-N}
\end{equation}
where $\kappa$ is now an arbitrary positive number.

Again we substitute $P_s = 2.5 \times 10^{-9}$, $\mathrm{Vol}(X_5)
= \pi ^3$, $v_2 = 4\pi$, $v_4 = \frac{8}{3}\pi ^2$,
$\frac{1}{3f_{\mathrm{NL}}}=0$. Furthermore, the minimum number of $e$-foldings
$\Delta \mathcal{N} =1$ and the lower limit value for the
tensor-to-scalar ratio during relativistic inflation $r=0.002$ from
(\ref{lowerbound}) is substituted in order to obtain a conservative
bound for $N$. For $g_s$, we take the value 0.1. 

\begin{itemize}
 \item D3-brane ($n=0$)\\
       In Sec.~\ref{sec:rev}, it was already shown that DBI inflation
       driven by a D$3$-brane is inconsistent with cosmological
       observations if $|f_{\mathrm{NL}}|$ is large. Nonetheless, as a 
       consistency check, let us show it again. The bound (\ref{G-N})
       turns out to be 
       $N>2.0\times 10^8\times (3.9\times 10^{17})^{1/\kappa}$. Taking
       the limit $\kappa\to 0+$, this lower bound for $N$ diverges,
       which implies that an infinitely large $N$ is required for a
       D3-brane to cause DBI inflation, i.e. DBI inflation is
       incompatible with WMAP3 data.\footnote{It was already
       mentioned in \cite{Alishahiha:2004eh} that the original DBI
       scenario with D3-branes required a large number of background
       charge $N\gtrsim 10^{10}$. This result
       was derived under the assumption of the inflaton potential
       consisting only of a quadratic term. We have shown here without
       specifying the form of the potential that a more stringent
       condition $N > \infty$ can be obtained by combining the
       results of Section~\ref{sec:rev}.} Hence we have obtained the same
       result, seen from a different perspective. 
 \item D5-brane ($n=1$)\\
       The bound (\ref{G-N}) in this case is 
       $N>6.8\times 10^5\times (7.0\times 10^{-3})^{1/\kappa}$. Since
       the number in parentheses is less than one, let us take the limit
       $\kappa\to\infty$. The lower bound for $N$ turns out to be 
       $6.8\times 10^5$. This exceeds the known maximal value of $N$ 
       (\ref{euler}) by an order.
 \item D7-brane ($n=2$)\\
       In this case we obtain 
       $N>9.7\times 10^4\times (1.2\times 10^{-9})^{1/\kappa}$ from
       (\ref{G-N}). The lower bound for $N$ becomes $9.7\times 10^4$ in
       the limit $\kappa\to\infty$. This still exceeds the maximal known
       value of $N$. However, adopting different values to the
       parameters may relax the lower bound. If the string coupling is 
       larger than about 0.2 (while keeping the other parameters fixed
       to the values discussed above), then the lower bound for $N$
       becomes compatible with (\ref{euler}). The same could be done by
       considering a throat with $\mathrm{Vol}(X_5) \lesssim 20$.
\end{itemize}

Note that the upper bound for $r$ (\ref{6}) and the lower bound for $N$
(\ref{G-N}) relax with a larger string coupling. The results above imply
the difficulty of maintaining perturbative control $g_s <1$ and
satisfying the upper bound for $N$ (\ref{euler}) at the same time.

\subsection{Upper Bound of $r$ in $B_{kl}$ Dominant Region}

We now consider the opposite extreme case, where $B_{kl}$ dominates
over $G_{kl}$. The results we obtain in this subsection is expected to
be relevant if perturbations of the 
CMB scale are generated in the large $\rho$ region (see 
Appendix~\ref{sec:Gregion}.).

In this subsection we derive an upper bound of $r$. In this sense the
analysis in this subsection is a counterpart to that of
subsection~\ref{subsec:rmax-Gregion}, where we have 
derived an upper bound of $r$ in the $G_{kl}$ dominant region. However,
the difference is that the analysis in 
this subsection requires (reasonable but explicit) assumptions about the
properties of the warp factor $h(\rho)$ and the NS-NS flux $B(\rho)$
which were reviewed in the beginning of this section, while the analyses in 
subsection~\ref{subsec:rmax-Gregion} were independent of those
properties.

In the next subsection we shall reinterpret the result of this
subsection as an lower bound of the background charge $N$.

Now let us start the analysis by introducing
%
\begin{equation}
 b_{2n} \equiv \int d^{2n}\xi \sqrt{\det\mathcal{B}_{kl}}.
\end{equation}
Based on examples of specific cycles in the KS solution, we expect that 
$b_2\sim v_2\sim 4\pi$ and that $b_4\sim v_4\sim\frac{8}{3}\pi^2$. 
Then (\ref{dphi}) and (\ref{T}) now take the form
%
\begin{eqnarray}
 d\phi & = & T_{3+2n}^{1/2}b_{2n}^{1/2}
  \left\{ g_s M {\alpha'}
   \ln\left(\frac{\rho}{\rho_b}\right)\right\}^{n/2}d\rho, 
  \label{B-dphi}\\
  T(\phi) & = & T_{3+2n}b_{2n}h^4\left\{g_s M \alpha'
   \ln\left(\frac{\rho}{\rho_b}\right)\right\}^n. \label{B-T} 
\end{eqnarray}

From the Lyth Bound (\ref{eqn:Lyth-bound}) and (\ref{B-dphi}), 
%
\begin{equation}
 \Delta\rho \simeq
  \frac{\Delta \mathcal{N}}{2^{3/2} T_{3+2n}^{1/2} 
  (g_s M\alpha')^{n/2}b_{2n}^{1/2}} 
  \frac{M_p r^{1/2}}
  {\left\{\ln\left(\frac{\rho_*}{\rho_b}
	     \right)\right\}^{n/2}}. \label{B3}
\end{equation}

Combining (\ref{TrPs}) with (\ref{B-T}),
%
\begin{equation}
 r=\frac{4\, T_{3+2n}^{1/2}\, b_{2n}^{1/2}\, (g_s M \alpha' )^{n/2}}
  {\pi\, P_s^{1/2}\, \left(1+\frac{1}{3f_{\mathrm{NL}}}\right)^{1/2}}
  \left(\frac{h_*}{M_p}\right)^2 
  \left\{\ln\left(\frac{\rho_*}{\rho_b}\right)\right\}^{n/2}. 
  \label{B2}
\end{equation}

From (\ref{i}) and (\ref{D3-1}), and again introducing an arbitrary
parameter $\kappa$ as in subsection~\ref{subsec:rmax-Gregion},
%
\begin{equation}
 \frac{1}{M_p^2} <
  \frac{(2\pi )^7\, g_s^2\, {\alpha'}^4}
  {2\, \mathrm{Vol}(X_5)} \, 
  \frac{h_*^4\, \rho_*^{\kappa}}{\Delta\rho}
  \left(\frac{1}{\rho_*}\right)^{\kappa+5}
  < 
  \frac{(2\pi )^7\, g_s^2\, {\alpha'}^4}{2\, \mathrm{Vol}(X_5)} \, 
  \frac{h_*^4 \, \rho_*^{\kappa}}{\Delta\rho}
  \left(\frac{1}{\Delta \rho}\right)^{\kappa +5} .\label{B5}
\end{equation}
Now $\kappa$ has to satisfy $\kappa > -5$. 
$\Delta \rho$ can be cancelled out from the far right hand side of
(\ref{B5}) with the use of (\ref{B3}).
Then, after cancelling out $M_p$ with the help of
(\ref{B2}), one can deduce an upper bound for $r$,
%
\begin{equation}
 r < \frac{(2\pi )^7\, g_s^2\, \alpha '^4}{2\, \mathrm{Vol}(X_5)}
\Biggl(\frac{\pi ^{1/2} \{ P_s (1+\frac{1}{3
  f_{\mathrm{NL}}})\}^{1/4}}{2 T_{3+2n}^{1/4} 
  b_{2n}^{1/4}(g_s M \alpha ')^{n/4}} 
\Biggr)^{\kappa+4}
\Biggl(\frac{\Delta \mathcal{N}}{2^{3/2} T_{3+2n}^{1/2} (g_s M
  \alpha ')^{n/2}b_{2n}^{1/2}} 
\Biggr)^{-\kappa -6}
\biggl(\frac{\rho_*}{h_*}\biggr)^{\kappa}
\biggl\{ \ln
\biggl(\frac{\rho_*}{\rho_b}\biggr)\biggr\}^{\frac{n}{4}(\kappa +8)}.
\label{r}
\end{equation}

Since $\frac{\rho}{h}=R$, $\rho \sim R$ is roughly the place where
the throat connects to the bulk (to be precise, one should be aware that 
the geometry deviates from the AdS geometry in the UV region, due to the
connection of the throat to the bulk). Since we consider the case of
inflation occurring within a single throat, $\rho_* < R$. Together with
(\ref{R}), (\ref{r}) can be rewritten in the following form,
%
\begin{equation}
 r < \frac{(2\pi )^7\, g_s^2\, \alpha '^4}{2\, \mathrm{Vol}(X_5)}
\Biggl(\frac{\pi ^{1/2} \{ P_s (1+\frac{1}{3
  f_{\mathrm{NL}}})\}^{1/4}}{2 T_{3+2n}^{1/4} (g_s M \alpha ')^{n/4}
  b_{2n}^{1/4}} 
\Biggr)^{\kappa+4}
\Biggl(\frac{\Delta \mathcal{N}}{2^{3/2} T_{3+2n}^{1/2} (g_s M
  \alpha ')^{n/2}b_{2n}^{1/2}} 
\Biggr)^{-\kappa -6}
R^{\kappa}\biggl\{ \ln
\biggl(\frac{R}{\rho_b}\biggr)\biggr\}^{\frac{n}{4}(\kappa +8)} .
\label{rr}
\end{equation}

From (\ref{R}) and (\ref{rho-b})
\begin{equation}
 \frac{R}{\rho_b} \sim \frac{2^{1/2}\, \pi}
  {\mathrm{Vol}(X_5)^{1/4}} \biggl(\frac{K}{g_s\,
  M}\biggr)^{1/4}\exp\biggl(\frac{2\pi K}{3g_s M}\biggr). 
\end{equation}
Since the exponential factor in the right hand side is approximately the
inverse of the warping at the tip of the throat (\ref{h0}), $\frac{2\pi
K}{3 g_s M}$ is expected to be larger than 1. Therefore,
\begin{equation}
 \ln \frac{R}{\rho_b}\sim \frac{2 \pi K}{3 g_s M}. \label{Rrhob}
\end{equation}

Substituting (\ref{Rrhob}) to (\ref{rr}) and taking $\kappa$ to 0
yields the following upper limit for $r$:
\begin{eqnarray}
  \mathrm{(D5)} & \quad & 
   r<\frac{2^3\, \pi\, b_2^2\, K^2\, 
   P_s(1+\frac{1}{3f_{\mathrm{NL}}})}{3^2\, \mathrm{Vol}(X_5)\, (\Delta
   \mathcal{N})^6} \simeq 3.5 \times 10^{-8}K^2 , \label{55}\\
 \mathrm{(D7)} & \quad &
  r < \frac{2\, b_4^2\, K^4\, 
  P_s(1+\frac{1}{3f_{\mathrm{NL}}})}{3^4\, \pi\, \mathrm{Vol}(X_5)\, (\Delta
  \mathcal{N})^6} \simeq 4.4 \times 10^{-10}K^4 . \label{56}
\end{eqnarray}
Here, $P_s = 2.5 \times 10^{-9}$, $\mathrm{Vol}(X_5)=\pi ^3$, 
$b_2 = 4\pi$, $b_4 = \frac{8}{3}\pi ^2$, $\frac{1}{3f_{\mathrm{NL}}}=0$,
and $\Delta \mathcal{N} =1$ have been substituted for the estimate of
the far right hand sides.

For these upper bounds to be compatible with the lower bound
(\ref{lowerbound}), $K$ needs to be larger than about 240 for a D5- and 46
for a D7-brane, which can readily be achieved.

\subsection{Lower Bound of $N$ in $B_{kl}$ Dominant Region}

The upper bound of $r$ (\ref{rr}) together with (\ref{Rrhob}) can be
transformed into a lower bound for $N$ with $\kappa > 0$: 
\begin{equation}
N > 
\frac{2^{n-1} 3^n \mathrm{Vol}(X_5) (\Delta \mathcal{N})^4}{\pi
  ^{3-n} b_{2n} K^n  P_s (1+\frac{1}{3 f_{\mathrm{NL}}})}
\Biggl(
\frac{2^{2n-5} \, 3^{2n}\, \pi ^{2n-3} \, \mathrm{Vol}(X_5)\, (\Delta
\mathcal{N})^6\,  r}
{b_{2n}^2 \,  P_s (1+\frac{1}{3 f_{\mathrm{NL}}}) K^{2n}}
\Biggr)^{\frac{4}{\kappa}} \label{B-N}
\end{equation}

Substituting the same values as above and $r=0.002$, the following
results can be obtained: 
\begin{itemize}
 \item D5-brane ($n=1$)\\
$NK > 3.0 \times 10^8 \times (240/K)^{8/\kappa}$. When
       $K\gtrsim 240$, taking $\kappa \to \infty$ yields a lower bound
       for the combination of the charge numbers, $NK > 3.0\times
       10^8$. For example, when the background charge takes the maximum
       value $N=75852$, the lower bound requires $K \gtrsim 4000$, which in
       this case is equivalent to $M \lesssim 20$. The smallness of $M$
       may invalidate the supergravity approximation in the warped throat.
 \item D7-brane ($n=2$)\\
$NK^2 > 2.7 \times 10^9 \times (46/K)^{16/\kappa}$. If $K
       \gtrsim 46$, then taking $\kappa \to \infty$ yields $NK^2 > 2.7 \times
       10^9$. When $N=75852$, the constraint turns out to be $K \gtrsim
       190$ ($M\lesssim 4000$).  
\end{itemize}

In terms of the background charge, a D7-brane doesn't seem to have
problems exciting DBI inflation. We also remark that the upper bound for
$r$ (\ref{55}) and (\ref{56}), and the lower bound for $N$ (\ref{B-N})
do not depend on the string coupling directly, in contrast to the bounds
derived in the $G_{kl}$ dominant region.

\section{Conclusion}
\label{sec:conc}

We have presented constraints on DBI inflation with large
non-Gaussianity, caused by a D-brane in relativistic motion towards the
tip of the throat. We focused on D5- and
D7-branes which wrap cycles of the warped throat. As expected, the upper
bound on the gravitational wave spectrum for D3-branes is relaxed for
wrapped D5- and D7-branes, due to the difference in the normalization
factor of the inflaton. However, for when the known maximal value for the Euler
number of the Calabi--Yau four-fold is considered, we showed the difficulty
in obtaining a sufficient number of background charge for producing the 
cosmological observables consistent with the WMAP3 data. The results of
this paper are insensitive to the details of the inflaton potential, and
could be applied to generic warped throats.

Our estimation imposes severe constraints on a D5-brane turning on DBI
inflation. However, for the case of a D7-brane producing the CMB scale
in the $B_{kl}$ dominant region of the throat, the constraint is
loosened. D7-brane DBI inflation in the $G_{kl}$ dominant region may
also be compatible with the Euler number bound, provided the string
coupling and the parameters of the throat are tuned to some appropriate
range. These cases may be the area to seek for brane 
inflation models generating observable signals in the sky. 

Our aim in this paper was to estimate whether DBI inflation can
reproduce observable signatures compatible with WMAP3 data in the
currently known construction of string theory. The results in this paper
indicate the difficulty of obtaining workable models due to the Euler
number bound. However, let us mention some possible scenarios in which
the bound may be relaxed. Though it is presently unknown whether those
possibilities can be implemented in string theory, they are worth
pointing out. One way of alleviating the Euler number bound is to
consider a throat with a base space that is
orbifolded~\cite{Alishahiha:2004eh,Becker:2007ui}. Orbifolding implies a
smaller volume of the base space $\mathrm{Vol}(X_5)$, leading to a
relaxation of the lower bound on $N$ ((\ref{G-N}) and (\ref{B-N})). If
the orbifolded base space decreases the volume of the wrapped cycles
(which works in the direction of tightening the bound), one can also consider
D-branes winding around the throat more than once to cancel the decrease
of $v_{2n}$. However, it is obvious from the results in this paper that
a substantial number of orbifolding (and winding number) is needed in
any case. At present it is far from clear whether this could be
justified in a consistent compactification. Other ways to relax the
Euler number bound are to add negative charges by considering
orientifolds or gluing on a whole throat of negative D3
charge~\cite{Douglas:2007tu} (though this would lead to an increase of
the compactified volume).\footnote{We thank Melanie Becker and Sarah
Shandera for pointing these out to us.} Finally, if the Euler number of
Calabi--Yau four-folds exceeding the known maximal value is to be found,
then of course the constraints derived in this paper will be
relaxed. 

An alternative approach to bypassing the background flux bounds is to
combine DBI inflation with other scenarios such as modular inflation
models. Then DBI inflation could generate large non-Gaussianity while
some other sector is producing enough $e$-foldings.

We should also mention backreaction of the mobile D-branes on the
background geometry. We treated wrapped D-branes as a probe, but since
D5- and D7-branes have a more significant backreaction than D3-branes
(especially when the winding number is large),
this potential problem deserves careful consideration. An order of
magnitude estimate of the backreaction by multi-winding D5-branes around
orbifolded base space is given in \cite{Becker:2007ui}.

We did not fix the form of the inflaton potential throughout this
paper. However, as we mentioned in subsection~\ref{subsec:diff}, it
should be noted that under some explicit form of potentials, even 
more severe constraints on the background charge could be obtained (see for
e.g. \cite{Baumann:2006cd,Becker:2007ui} where a tight upper bound on
$N$ is derived under the assumption of a quadratic inflaton
potential). The constraints we investigated are conservative bounds,
which can be applied to general situations.

\begin{acknowledgments}
 T.K. and S.K. would like to thank Katsuhiko Sato for his encouragement
 and support. S.K.'s work was in part supported by JSPS through a
 Grant-in-Aid for JSPS Fellows. S.M. was supported by
 MEXT through a Grant-in-Aid for Young Scientists (B) No.~17740134, and
 by JSPS through a Grant-in-Aid for Creative Scientific Research
 No.~19GS0219 and through a Grant-in-Aid for Scientific Research (B)
 No.~19340054. 
\end{acknowledgments}

\appendix

\section{Klebanov--Strassler Solution}
\label{sec:KS}

The KS geometry \cite{Klebanov:2000hb,Herzog:2001xk} is
\begin{equation}
 ds^2 = \tilde{h}^{-1/2}(\tau) \eta _{\mu\nu} dx^{\mu}dx^{\nu} +
  \tilde{h}^{1/2}(\tau)ds_6^2  \label{KSmetric}
\end{equation}
where $x^{\mu}$ ($\mu = 0,\cdots 3$) are 4-dimensional external
coordinates and $ds_6^2$ is the metric of the deformed
conifold~\cite{CommentsonConifolds}  
\begin{equation}
 ds_6^2 = \frac{\epsilon ^{4/3}}{2}K(\tau) \biggl[ \frac{1}{3 K(\tau
  )^3}\Bigl(d\tau^2 + (g^5)^2\Bigr) + \cosh ^2 \Bigl(\frac{\tau}{2}\Bigr)
\Bigl((g^3)^2+(g^4)^2\Bigr) + \sinh^2 \Bigl(\frac{\tau}{2}\Bigr) \Bigl(
(g^1)^2+(g^2)^2\Bigr) 
\biggr] \label{KSds6}
\end{equation}
where
\begin{equation}
 K(\tau) = \frac{(\sinh(2 \tau)-2\tau)^{1/3}}{2^{1/3} \sinh \tau}
\end{equation}
and $g^i$ ($i=1,\cdots 5$) are orthonormal basis \cite{Minasian:1999tt}
defined by 
\begin{equation}
 g^1=\frac{e^1-e^3}{\sqrt{2}},\ \ g^2=\frac{e^2-e^4}{\sqrt{2}},\ \ 
g^3=\frac{e^1+e^3}{\sqrt{2}},\ \ g^4=\frac{e^2+e^4}{\sqrt{2}},\ \ 
g^5=e^5
\end{equation}
with
\begin{eqnarray}
 e^1 = -\sin\theta_1\, d\phi_1, \ \ \ \ \ 
 e^2 = d\theta_1, \ \ \ \ \ 
 e^3 = \cos\psi \sin \theta_2\, d\phi_2 - \sin\psi\, d\theta_2,
  \nonumber \\
 e^4 = \sin\psi \sin \theta_2\, d\phi_2 + \cos\psi\, d\theta_2, \ \ \ \
  \ 
 e^5 = d\psi + \cos\theta_1\, d\phi_1 + \cos\theta_2\, d\phi_2.
\end{eqnarray}

Away from the tip, the geometry is approximately $\mathrm{AdS}_5 \times S^5$.

The base of the cone has the topology of $S^2 \times S^3$. As one
approaches the tip, the radius of $S^2$ shrinks to zero, while the
radius of $S^3$ remains finite. Hence, the geometry is roughly $S^3
\times R^3$ at the tip of the throat. The $S^3$ subspace is a 3-cycle, which
is referred to as the A-cycle. Another dual 3-cycle which is the $S^2$
times a circle extending along the radial direction is called the
B-cycle. The R-R 3-form flux $F_3$ and NS-NS 3-form flux $H_3$ is
supported on these cycles, whose quantization conditions are
\begin{equation}
 \frac{1}{2\pi \alpha '}\int _A F_3 = 2\pi M, \ \ \ 
\frac{1}{2 \pi \alpha '}\int _B H_3 = -2\pi K, \label{MK}
\end{equation}
where $M$ and $K$ are integers, and 
\begin{equation}
 g_s^2 F_3^2 = H_3^2. \label{77}
\end{equation}
The tadpole condition requires
\begin{equation}
 KM = \frac{\chi}{24} \label{tadpole}
\end{equation}
where $\chi$ is the Euler number of a Calabi--Yau four-fold.

$F_3$ and $B_2$ have the $Z_2$ symmetric ($(\theta_1,\ \phi_1)
\leftrightarrow (\theta_2,\ \phi_2)$) ansatz:
\begin{equation}
 F_3 = \frac{M \alpha '}{2}\bigl[ g^5 \wedge g^3 \wedge g^4 + 
d\{F(\tau) (g^1 \wedge g^3 + g^2 \wedge g^4)\}\bigr],
\end{equation}
\begin{equation}
 B_2 = \frac{g_s M \alpha '}{2}\bigl[f(\tau) g^1 \wedge g^2 + k(\tau)
  g^3 \wedge g^4 \bigr]. \label{KSB2}
\end{equation}
Combining with (\ref{77}), the dilaton $\Phi$ and the R-R scalar $C_0$
can consistently be set to zero. The BPS saturated solution found by
Klebanov and Strassler is
\begin{equation}
 F(\tau) = \frac{\sinh \tau - \tau}{2 \sinh \tau}, \ \ \ 
 f(\tau) = \frac{\tau \coth \tau -1}{2 \sinh \tau}(\cosh \tau -1), \ \ \ 
 k(\tau) = \frac{\tau \coth \tau -1}{2 \sinh \tau}(\cosh \tau +1),
\end{equation}
and
\begin{equation}
 \tilde{h}(\tau) = 2^{2/3}(g_s M \alpha ')^2 \epsilon ^{-8/3} I(\tau),
  \label{htau} 
\end{equation}
where
\begin{equation}
 I(\tau)  = \int ^{\infty}_{\tau} dx \frac{x \coth x-1}{\sinh ^2
  x}(\sinh (2x)-2x)^{1/3}. \label{Itau}
\end{equation}
For this solution,
\begin{equation}
 C_4 = g_s^{-1}\tilde{h}^{-1} dx^0 \wedge dx^1 \wedge dx^2 \wedge dx^3
\end{equation}
in a particular gauge. For large $g_s M$ the curvature is small
everywhere and we can trust the supergravity description.

\vspace{\baselineskip}
$I(\tau)$ reaches a finite value as $\tau$ approaches zero,
\begin{equation}
 I(0) = a_0\ \  \mathrm{with}\ \  a_0 \simeq 0.71805. \label{azero}
\end{equation}
Since the warp factor at the tip of the throat can be characterized
by the flux integer numbers $M$ and $K$ introduced in (\ref{MK}),
\begin{equation}
 \tilde{h}^{-1/4}(0) \simeq \exp \biggl(-\frac{2 \pi K}{3 g_s M}\biggr),
  \label{htip} 
\end{equation}
the deformation parameter $\epsilon$ in (\ref{htau}) is
\begin{equation}
 \epsilon \simeq 2^{1/4} a_0^{3/8} (g_s M \alpha')^{3/4} \exp 
\biggl(-\frac{\pi K}{g_s M}\biggr). \label{epsilon}
\end{equation}

The correspondence between the KS metric (\ref{KSmetric}) and the
generic metric (\ref{metric}) we use in this paper is clear by the
change of variables
\begin{equation}
 d\rho = \frac{\epsilon ^{2/3}}{\sqrt{6} K(\tau )}d\tau, \label{rhotau}
\end{equation}
and 
\begin{equation}
 h(\rho) = \tilde{h}(\tau)^{-1/4}. \label{htildeh}
\end{equation}

In the large $\tau$ region, the metric of the deformed
metric (\ref{KSds6}), the relation between $\rho$ and $\tau$ (\ref{rhotau}),
and the NS-NS 2-form potential $B_2$ (\ref{KSB2}) is
\begin{equation}
 ds_6^2 \simeq d\rho^2 + \rho ^2 \biggl\{
\frac{1}{6}\bigl( (g^1)^2 + (g^2)^2 +(g^3)^2+(g^4)^2\bigr)
+ \frac{1}{9}(g^5)^2
\biggr\}, \label{large-ds6}
\end{equation}
\begin{equation}
 \rho \simeq \frac{3^{1/2}}{2^{5/6}}\epsilon ^{2/3} e^{\tau/3},
  \label{large-thotau}
\end{equation}
\begin{equation}
 B_2 \simeq \frac{3}{4}g_s M \alpha ' \ln \biggl(\frac{\rho}{\rho
  _b}\biggr) [g^1 \wedge g^2 - g^3 \wedge g^4], \label{large-B2}
\end{equation}
where
\begin{equation}
 \rho _b \equiv \frac{3^{1/2} e^{1/3}}{2^{5/6}}\epsilon ^{2/3} 
\simeq \frac{3^{1/2} e^{1/3} a_0^{1/4}}{2^{2/3}} (g_s M \alpha ')^{1/2}
\exp \biggl( - \frac{2 \pi K}{3 g_s M}\biggr). \label{rhob}
\end{equation}
The far right hand side of (\ref{rhob}) is from (\ref{epsilon}).

In the small $\tau$ region ($\tau \lesssim 1$),
\begin{equation}
 ds_6^2 \simeq d\rho^2 +
  \rho^2\biggl\{\frac{1}{2}(g^1)^2 + \frac{1}{2}(g^2)^2\biggr\} +
  \rho_0^2\biggl\{(g^3)^2+(g^4)^2+\frac{1}{2}(g^5)^2\biggr\} 
 \label{small-ds6}
\end{equation}
where
\begin{equation}
 \rho_0 \equiv \frac{\epsilon^{2/3}}{2^{1/3}3^{1/6}}, \label{rho0}
\end{equation}
\begin{equation}
 \rho = \frac{\epsilon ^{2/3}\tau}{2^{5/6}3^{1/6}}, \label{small-rhotau}
\end{equation}
\begin{equation}
B_2 \simeq \frac{g_s M \alpha '}{2}\biggl[\frac{\tau^3}{12} (g^1 \wedge
 g^2) + \frac{\tau}{3} (g^3 \wedge g^4)\biggr]. \label{small-B2}  
\end{equation}

\section{DBI Inflation Near the Tip of the Throat}
\label{sec:tip}

In this appendix, we discuss DBI inflation in the non-AdS region, nearby
the tip of the throat. We take the KS throat as an example and show
that the CMB scale is produced away from this region.

In a KS throat, the conifold singularity is smoothed out by turning on
background flux. While the $S^2$ of the base space disappears as one
approaches the tip of the deformed conifold, the $S^3$ remains finite
and the warp factor approaches a constant value (\ref{htip}).

In the region $\tau \lesssim 1$ (i.e. $\rho \lesssim 0.8 \rho_0$, where
$\rho_0$ is defined in (\ref{rho0})), $\tau$ and $\rho$ become
proportional to each other, as can be seen from
(\ref{small-rhotau}). The metric and the NS-NS 2-form potential $B_2$ in
the region is given by (\ref{small-ds6}) and (\ref{small-B2}). 

In the $g^1$ and $g^2$ directions,
the behavior of the metric and $B_2$ in this region are respectively
\begin{equation}
 \frac{\rho^2}{h^2} \propto \tau^2,\ \ \ f(\tau) \propto \tau^3.
\end{equation}
In the $g^3$ and $g^4$ directions,
\begin{equation}
 \frac{\rho_0^2}{h^2} \propto \tau^0,\ \ \ k(\tau) \propto \tau.
\end{equation}
Furthermore, $B_2$ does not have a leg in the $g^5$ direction.

Hence it is expected that the $B_2$ term is negligible compared to the
term coming from the metric in the DBI action
(\ref{DBIaction}). Therefore we ignore $B_2$ in estimating possible
$e$-fold numbers near the tip.

Ignoring $B_2$, the applicability of the lower bound for the tensor to
scalar ratio (\ref{lowerbound}) to the region near the tip with the
metric (\ref{small-ds6}) can be verified in a similar way to the
discussion in subsection~\ref{subsec:gelobr}. Taking the angular brane
coordinates to diagonalize $G_{kl}$, it is clear that $\partial (h^2
G_{kk})/ \partial \rho \ge 0$. Hence (\ref{ddrho}) is verified, leading
to the lower bound (\ref{lowerbound}).

Also, the equations introduced in subsection~\ref{subsec:rmax-Gregion} to
estimate the upper bound for $r$ in the $G_{kl}$ dominant region can be
used here likewise. However, it should be noted that the constants
$\mathrm{Vol}(X_5)$ and $v_{2n}$ now become dependent on $\rho$, due to
the $S^2$'s constant radius. The effective unit-radius dimensionless
volume of the wrapped $2n$-cycle is now defined as
\begin{equation}
 v_{2n}(\rho) \equiv \biggl(\frac{h}{\rho}\biggr)^{2n} \int d^{2n}\xi
  \sqrt{\mathrm{det}(G_{kl})}. \label{vrho}
\end{equation}

Now let us give a rough estimate of the number of $e$-foldings which can
be generated in the region near the tip. We will see that the $e$-fold
number is suppressed to a negligible amount by the warping of the
throat. 

A typical element of $G_{kl}$ takes the form of a quadratic equation of
$\rho$ or $\rho_0$, divided by $h^2$. Therefore, $v_{2n}(\rho)$ can
practically be expressed as a product of $\rho^{-2n}$ and a $2n$-order
polynomial of $\rho$ or $\rho_0$. Therefore, in the ``tip'' region, i.e.
$\rho \lesssim 0.8\rho_0$, the effective unit-radius dimensionless
volume is  
\begin{equation}
 v_{2n}(\rho) \sim \frac{\mathcal{O}\bigl(\rho_0^{2n}\bigr)}{\rho
  ^{2n}}. \label{vpoly}
\end{equation}

Combining the Lyth Bound (\ref{eqn:Lyth-bound}), (\ref{G-dphi}), and
(\ref{vpoly}), we obtain
\begin{equation}
 d\mathcal{N} \sim \frac{T_{3+2n}^{1/2}\, \mathcal{O}\bigl(
  \rho_0^{n}\bigr)}{M_p\, r^{1/2}\, h^n} d\rho. \label{roughN}
\end{equation}
The number of $e$-foldings can be evaluated by integrating the right
hand side. We estimate the maximal number of $e$-foldings by fixing the
tensor to scalar ratio to the smallest $r=r_{\mathrm{min}}$ in the
region of integration. Similarly, we fix the warp factor to the value at
the tip of the throat $h=h_{\mathrm{tip}}$. Integrating between 0 and
$\rho_0$, (it should be noted that inflation actually ends before
reaching $\rho=0$ in most models)
\begin{equation}
 \mathcal{N}_{\mathrm{tip}} \lesssim \frac{T_{3+2n}^{\frac{1}{2}}\,
  \mathcal{O}\bigl((\rho_0)^{n+1}\bigr)}{M_p\, r_{\mathrm{min}}^{\frac{1}{2}}\, 
  h_{\mathrm{tip}} ^n} \sim \frac{g_s^{\frac{n}{2}}\, M^{\frac{n+1}{2}}\,
  h_{\mathrm{tip}}}{\alpha '^{\frac{1}{2}}\, M_p\,
  r_{\mathrm{min}}^{\frac{1}{2}}}. 
 \label{Nnplus1} 
\end{equation}
In obtaining the far right hand side, we have used
\begin{equation}
 \rho_0 = \frac{a_0^{1/4}}{2^{1/6}\, 3^{1/6}} (g_s M \alpha
  ')^{1/2}h_{\mathrm{tip}} 
  \label{rho0h0}
\end{equation}
which is a combination of (\ref{htip}), (\ref{epsilon}), (\ref{htildeh}),
and (\ref{rho0}). Hence we have shown that the number of $e$-foldings
that can be generated in the ``tip'' region is suppressed by the warp
factor.  

\vspace{\baselineskip}
Now we confirm the above result with an explicit example. Let us
consider a D5-brane which wraps a 2-cycle specified by
\begin{equation}
 \psi =0, \ \theta_1 = \theta_2,\ \phi_1 = -\phi_2. \label{2-cycle}
\end{equation}
Then the following can be derived
\begin{equation}
\int d^2\xi \sqrt{\mathrm{det}(G_{lk}-B_{lk})} = \frac{2^3\, \pi\,
 \rho_0^2}{h^2} \biggl\{ 1+\biggl(\frac{h^2\, g_s M \alpha '
 \rho}{2^{2/3}\, 3\, \rho_0^3}\biggr)^2\biggr\}^{1/2}.
 \label{D5-BvsG}
\end{equation}
The first term in parentheses of the right hand side originates from
$G_{kl}$, and the second term from $B_{kl}$. Approximating the warp
factor by the value at the tip, then from (\ref{rho0h0}) the ratio
between them is 
\begin{equation}
 \frac{B\, \mathrm{term}}{G\, \mathrm{term}} \simeq 0.2\times \frac{\rho
  ^2}{\rho_0^2}. 
\end{equation}
Hence it is clear that $B_2$ can be ignored in the ``tip'' region (i.e. $\rho
\lesssim 0.8 \rho_0$).

Ignoring the $B_2$ term, the effective unit-radius dimensionless
volume is
\begin{equation}
 v_{2}(\rho) = \bigl(\frac{\rho_0}{\rho}\bigr)^2 8\pi . \label{v2rho}
\end{equation}  
This simple form originates from the fact that the 2-cycle 
specified by (\ref{2-cycle}) wraps only the non-vanishing $S^3$. 

The integration which led (\ref{roughN}) to (\ref{Nnplus1}) can be
explicitly carried out for the simple form of $v_2$ (\ref{v2rho}). 
Keeping also the numerical factors, we obtain
\begin{equation}
 \mathcal{N}_{\mathrm{tip}} \le \frac{0.007}{r_{\mathrm{min}}^{1/2}}
 \times \frac{g_s^{1/2}\, M}{\, M_p\, \alpha'^{1/2}}\times
 h_{\mathrm{tip}} 
 \label{D5-Ntip} 
\end{equation}
Assuming that the CMB scale is generated in the ``tip'' region, we
substitute $r_{\mathrm{min}}=0.002$ which is the lower bound of $r$
over the observable scales for DBI inflation (\ref{lowerbound}). Then 
the right hand side of (\ref{D5-Ntip}) is expected to be dominated by
the warping at the tip of the throat, suppressing
$\mathcal{N}_{\mathrm{tip}}$ to be much smaller than the
minimum number of $e$-foldings produced while the observable scales are
generated, i.e. $\Delta\mathcal{N}\simeq 1$.

Since the CMB scale was generated 30 to 60 $e$-foldings before
the end of inflation, the above estimates indicate that the CMB scale
cannot be produced in the non-AdS region nearby the tip of the KS
throat.

Throughout this paper we consider warped throats. However,
we should note that in the case of a barely warped  
throat (e.g. $h_{\mathrm{tip}}\sim \mathcal{O}(1)$), there will be
a large non-AdS region and it is possible that a
sufficiently large number of $e$-foldings will be generated in the
region. (In that case, the volume of the internal space must be large for
explaining the hierarchy of the universe.)

\section{Number of $e$-Foldings in the $G_{kl}$ Dominant Region}
\label{sec:Gregion}

In Sections~\ref{sec:disc} and \ref{sec:charge} we discussed constraints
on $r$ and $N$ for the case 
of the observed CMB scale being produced in the $G_{kl}$ or $B_{kl}$
dominant region. The analyses indicate that DBI inflation
requires a large Euler number of the Calabi--Yau four-fold, which exceeds
the maximal value (\ref{euler}) under typical values for various
parameters. However, for DBI inflation with a D7-brane in the $B_{kl}$
dominant region, the constraint is relaxed.

In this appendix, we roughly estimate the number of $e$-foldings than can be
produced in the $G_{kl}$ dominant region and consider the place where
the CMB scale is produced.

The typical ratio between the components of $G_{kl}$ and $B_{kl}$ is
\begin{equation}
 \frac{\mathrm{components\ of\ }G_{kl}}{\mathrm{components\ of\ }B_{kl}}
  = \frac{\rho^2}{h^2} \times 
  \biggl\{ g_s M \alpha ' \ln
  \Bigl(\frac{\rho}{\rho_b}\Bigr)\biggr\}^{-1} = 
  \frac{R^2}{g_s M \alpha ' \ln (\frac{\rho}{\rho_b})}. \label{G/B}
\end{equation}

Therefore, the $G_{kl}$ dominant region is roughly
\begin{equation}
 \rho < \rho_b \exp\biggl(\frac{R^2}{g_s M \alpha '}\biggr) \equiv \rho_{G/B}.
\end{equation}

From the Lyth Bound (\ref{eqn:Lyth-bound}), (\ref{R}), and (\ref{G-dphi}),
\begin{equation}
 d\mathcal{N} \simeq \frac{v_{2n}^{1/2}\,  N^{n/4}}{2^{n/2}\, \pi^{3/2}\,
  g_s^{(2-n)/4}\, \alpha'\,  M_p\,  \mathrm{Vol}(X_5)^{n/4}\, r^{1/2}}
  \, d\rho. \label{dom-Ndrho} 
\end{equation}

The number of $e$-foldings $\mathcal{N}_G$ generated in the $G_{kl}$
dominant AdS region 
can be obtained by integrating (\ref{dom-Ndrho}) between $\rho_{G/B}$
and the IR boundary of the AdS region, which we approximate to
0. Note that $r$ depends on $\rho$. Combining (\ref{G/B}) with (\ref{R})
and (\ref{rho-b}) yields the following, 
\begin{eqnarray}
 \lefteqn{
 \mathcal{N}_{G} \lesssim \frac{v_{2n}^{1/2}\,
  M^{(n+1)/2}}{2^{n/2}\, \pi^{3/2}\, \alpha '^{1/2}\, M_p\,
  r_{\mathrm{min}}^{1/2}}}\hspace{1cm}
  \nonumber \\  & & \times \Bigl(\frac{ g_s}{\mathrm{Vol}(X_5)}\Bigr)^{n/4}
 \Bigl(\frac{K}{M}\Bigr)^{n/4} \exp \Biggl[ \biggl\{ \frac{2
  \pi^2}{g_s^{1/2}\, \mathrm{Vol}(X_5)^{1/2}}-\frac{2\pi}{3
  g_s}\Bigl(\frac{K}{M}\Bigr)^{1/2}\biggr\}
  \Bigl(\frac{K}{M}\Bigr)^{1/2} \Biggr], \label{NG}
\end{eqnarray}
where we pulled $r$ out of the integration by introducing the
smallest $r=r_{\mathrm{min}}$ in the region of integration, thereby an
upper bound for $\mathcal{N}_G$ is obtained.

The second line of (\ref{NG}) is expressed as a function of $K/M$. It
increases monotonically when $K/M$ is small, and when 
\begin{equation}
 \frac{K}{M}= \frac{9g_s}{8}\Biggl\{\frac{\pi^2}{\mathrm{Vol}(X_5)} +
  \frac{n}{3\pi}+ \biggl(\frac{\pi^4}{\mathrm{Vol}(X_5)^2}+
  \frac{2n\pi}{3\mathrm{Vol}(X_5)}\biggr)^{1/2}\Biggr\},  \label{K/M}
\end{equation}
it reaches its maximum value
\begin{equation}
 \Biggl[\frac{3
  g_s}{4\pi}\Biggl\{\frac{\pi^2}{\mathrm{Vol}(X_5)}+
  \biggl(\frac{\pi^4}{\mathrm{Vol}(X_5)^2}
  +\frac{2n\pi}{3\mathrm{Vol}(X_5)}\biggr)^{1/2}\Biggr\}\Biggr]^{n/2} 
 \times \exp\Biggl[\frac{3\pi^3}{4\mathrm{Vol}(X_5)}-\frac{n}{4}+
  \frac{3\pi}{4} \biggl(\frac{\pi^4}{\mathrm{Vol}(X_5)^2}+
  \frac{2n\pi}{3\mathrm{Vol}(X_5)}\biggr)^{1/2}\Biggr], \label{max}
\end{equation}
and then it drops sharply as $K/M$ becomes larger than (\ref{K/M}). Note
that the maximum value (\ref{max}) is a monotonically increasing
function of $g_s$, and a monotonically decreasing function of
$\mathrm{Vol}(X_5)$. 

Assuming values such as $\mathrm{Vol}(X_5) = \pi^3$, $v_2=4\pi$, $v_4 =
\frac{8}{3}\pi^2$, $g_s = 0.1$, and $\alpha ' M_p^2=1000$, then from
(\ref{NG}), (\ref{K/M}), and (\ref{max}) the following upper bounds can
be obtained:
\begin{equation}
 \mathrm{(D5)}\ \ \ \ \mathcal{N}_{G} \lesssim 3\times 10^{-3}
  \frac{M}{r_{\mathrm{min}}^{1/2}}\biggl(\frac{K}{M}\biggr)^{1/4} \exp
  \Biggl[ \Biggl\{11-21\biggl(\frac{K}{M}\biggr)^{1/2}\Biggr\}
  \biggl(\frac{K}{M}\biggr)^{1/2}\Biggr] \le 7\times 10^{-3} \times
  \frac{M}{r_{\mathrm{min}}^{1/2}} ,\label{D5NG}
\end{equation}
\begin{equation}
 \mathrm{(D7)}\ \ \ \ \mathcal{N}_{G} \lesssim 8\times 10^{-4}
  \frac{M}{r_{\mathrm{min}}^{1/2}}\biggl(\frac{K}{M}\biggr)^{1/2} \exp
  \Biggl[ \Biggl\{11-21\biggl(\frac{K}{M}\biggr)^{1/2}\Biggr\}
  \biggl(\frac{K}{M}\biggr)^{1/2}\Biggr] \le 1\times 10^{-3} \times
  \frac{M}{r_{\mathrm{min}}^{1/2}} .\label{D7NG}
\end{equation}

The far right hand sides of (\ref{D5NG}) and (\ref{D7NG}) are obtained
when $K/M \sim 0.1$. For example, let us substitute $N=75852$,
$M=10K\simeq 870$, and 
$r_{\mathrm{min}}=0.002$, then the upper bounds for
$\mathcal{N}_G$ are about 140 for a D5- and 20 for a D7-brane. However,
we should remark that $r_{\mathrm{min}}$ could be smaller than
0.002. Perturbations at scales smaller than the CMB scale could be
generated by standard slow-roll inflation. Even if DBI inflation
continues until the end of inflation, if $1\!-\!n_s$ decreases at
scales smaller than the CMB scale, then the lower bound for $r$ in
(\ref{lowerbound}) becomes smaller than 0.002.

Since the upper bound for $\mathcal{N}_G$ varies with the parameters
(especially sensitive to $K/M$ and $\mathrm{Vol}(X_5)$), we end this
appendix by stating that the region where the CMB scale is generated
depends on the details of the DBI inflation model.

\section{Determinant of $G_{kl}-B_{kl}$}
\label{sec:det}

Taking a gauge in which the metric $G_{kl}$ is diagonalized,
the difference between the $n\times n$ $G_{kl}$ and an antisymmetric $B_{kl}$
takes the following form:
\begin{equation}
 G_{kl} - B_{kl} = \left(
\begin{array}{cccc}
G_{11} & B_{12} & \ldots & B_{1n} \\
-B_{12} & G_{22} & \ddots & \vdots \\
\vdots & \ddots & \ddots & B_{n\!-\!1 n} \\
-B_{1n} & \ldots & -B_{n\!-\!1 n} & G_{nn}
\end{array} \right).\label{matrix}
\end{equation}

The expansion of the determinant of (\ref{matrix}) in terms of the diagonal
components of $G_{kl}$ is
\begin{equation}
 \mathrm{det}(G_{kl}-B_{kl}) = \sum_{m=0}^{n} \sum_{G^{n\!-\!m}}
  G_{i_{m\!+\!1}i_{m\!+\!1}}\cdots G_{i_{n}i_{n}} \times \mathrm{det}\left(
\begin{array}{cccc}
0 & B_{i_1i_2} & \ldots & B_{i_1i_m} \\
-B_{i_1i_2} & 0 & \ddots & \vdots \\
\vdots & \ddots & \ddots & B_{i_{m\!-\!1}i_m} \\
-B_{i_1i_m} & \ldots & -B_{i_{m\!-\!1}i_m} & 0
\end{array} \right), \label{mat}
\end{equation}
where $\sum_{G^{n\!-\!m}}$ requires to sum up all the
combinations of choosing $(n\!-\!m)$ numbers of different $G_{kk}$.

The matrix on the right hand side of (\ref{mat}) is built from
$G_{kl}-B_{kl}$ by erasing the $i_{m\!+\!1},\, i_{m\!+\!2},\, \cdots,
i_{n}$th rows and $i_{m\!+\!1},\, i_{m\!+\!2},\, \cdots, i_{n}$th columns,
and changing the diagonal terms to zero. Since it is an $m\times m$
antisymmetric matrix, its determinant is zero for odd $m$. Hence 
$\mathrm{det}(G_{kl}-B_{kl})$ is expanded with even-ordered $G_{ll}$.

If $m$ is even, the determinant of an antisymmetric matrix is equal to
the square of the Pfaffian of the 
matrix, which is a polynomial in the components.
When $X$ is a $2p\times 2p$ antisymmetric matrix with components
$x_{ij}$, the Pfaffian of 
$X$ is
\begin{equation}
 \mathrm{Pf}(X) = \frac{1}{2^p p!}\sum_{\sigma \in
  S_{2p}}\mathrm{sgn}(\sigma) \prod_{i=1}^px_{\sigma(2i-1)\,
  \sigma(2i)},
\end{equation}
where $S_{2p}$ is the symmetric group. Then the
determinant of $X$ is 
\begin{equation}
 \mathrm{det}(X) = \mathrm{Pf}(X)^2.
\end{equation}


\end{document}